# Silicon Photonic Beam Steerer Based on Metalens Focal Plane Array

Chung-Yu Hsu,[1,†] Ping-Yen Hsieh,[1,†] Hsun-Sung Chiu,[1] Li-Jun Tung,[1] Chieh-Chih Yu,[1] Ko-Chi Chen,[1] Yu-Heng Hong,[2] Hao-Chung Kuo,[1,2] Chi-Wai Chow,[1] and You-Chia Chang[1,*]

[1]*Department of Photonics, College of Electrical and Computer Engineering, National Yang Ming Chiao Tung University, Hsinchu 30069, Taiwan*
[2]*Semiconductor Research Center, Hon Hai Research Institute, Taipei 11492, Taiwan*

*\*youchia@nycu.edu.tw*

**Abstract:** Focal plane arrays (FPAs) promise robust solid-state beam steering for LiDAR and free-space optical communications. However, the need for external collimation lenses hinders chip-scale compactness. Discrete switching between FPA elements further introduces blind spots and limits the number of resolvable points, restricting applications that require continuous tracking. Here, we demonstrate a silicon photonic beam steerer based on a metalens FPA that monolithically integrates the collimation lens on-chip. Thermo-optic prisms enable continuous fine-tuning, eliminating blind spots and tripling the number of resolvable points. Continuous steering over a 62° field of view is achieved while maintaining high beam quality, with an average sidelobe suppression ratio of 19 dB.

## 1. Introduction

Beam steerers are key components in scanning-based technologies [1], such as light detection and ranging (LiDAR) [2–6], free-space optical communications (FSO) [7,8], and laser-scanning microscopy [9,10]. LiDAR enables 3D mapping of the physical world, holding great promise for autonomous vehicles, robotics, and the Internet of Things. With the rapid advancement of artificial intelligence (AI), access to high-quality 3D spatial information has become essential for training AI models and for enhancing the interaction between digital and physical domains. FSO can further establish highly flexible communications between spatially distributed nodes, including dynamic airborne platforms such as unmanned aerial vehicles (UAVs) [11]. These trends drive demand for ultracompact, highly reliable LiDAR and FSO systems. However, conventional mechanical beam steerers are too bulky to meet this demand, and their long-term reliability is constrained by the moving parts that undergo repeated mechanical motion. Silicon photonics offers a solid-state solution that miniaturizes a beam steerer to the chip scale with no moving parts. Thanks to its compatibility with the complementary metal-oxide-semiconductor (CMOS) processes, silicon photonics enables high-yield integration of a large number of diverse components on the same chip. The ability to scale up the number of components in a circuit is particularly attractive, as the resolution of beam steering depends critically on this number [12].

Optical phased arrays (OPAs) are among the most extensively explored beam-steering technologies on the silicon photonic platform [3,6–10,13–20]. In an OPA, light is distributed to multiple emitters, each with a controllable phase, to synthesize a steerable beam. Wide field-of-view (FOV) steering up to 180° [16,17] and large-scale integration with 9,216 phase shifters have been demonstrated [18]. However, due to fabrication-induced phase variations, phase calibration for a large number of phase shifters, typically performed using iterative algorithms or training, has been time-consuming and unreliable [19]. For instance, calibrating a 128-

channel OPA can take ~100 hours for data acquisition and training [19], and real-time on-chip phase monitoring remains challenging [20].

In contrast, focal plane arrays (FPAs) provide a more robust beam-steering mechanism [2,12,21–25], as phase calibration is not required. An FPA comprises an optical switch network, an array of emitters, and a collimation lens. The optical switch network routes light to a selected emitter, while the lens—positioned at a focal length *f* from the emitter array—produces a collimated output beam. The beam direction is determined by the selected emitter location. Switch calibration is straightforward and relies solely on power monitoring. However, the discrete switching operation inherently limits steering to discrete angles, resulting in blind spots within the FOV. The discrete nature severely limits its applicability to FSO systems, as link establishment requires fine alignment and tracking between transmitters and receivers. Moreover, the external collimation lens must be precisely aligned and packaged with the silicon photonic chip, adding system-level bulk and increasing packaging cost [2,5,12,21].

Another common limitation shared by both OPA and FPA technologies is that the number of resolvable points is approximately equal to the number of array elements *N* [12]. Consequently, achieving a practically useful high-resolution beam steerer requires implementing a large-scale array, which occupies significant chip area and complicates the control circuitry. This limitation can be mitigated by adopting irregularly spaced OPAs. For instance, employing a Costas array can enhance the number of resolvable points to $N^2$ [26]. However, this enhancement comes at the expense of beam quality and efficiency, as sidelobe energy is redistributed across a broad angular range, resulting in elevated background noise.

In this work, we present a solution to these long-standing issues of FPA: we achieve continuous beam steering without blind spots and monolithically integrate the collimation lens on-chip using metasurface technology [27–29]. A metalens FPA provides channel-dependent discrete coarse steering, which is combined with a continuous fine-tuning mechanism based on thermo-optic (TO) prisms. This architecture increases the number of resolvable points by a factor of three, yielding *3N* points using only *N* waveguide channels without compromising beam quality. We experimentally demonstrate continuous steering over a FOV of 62°, while achieving an average sidelobe suppression ratio (SLSR) of 19 dB.

## 2. Operation principle and simulation

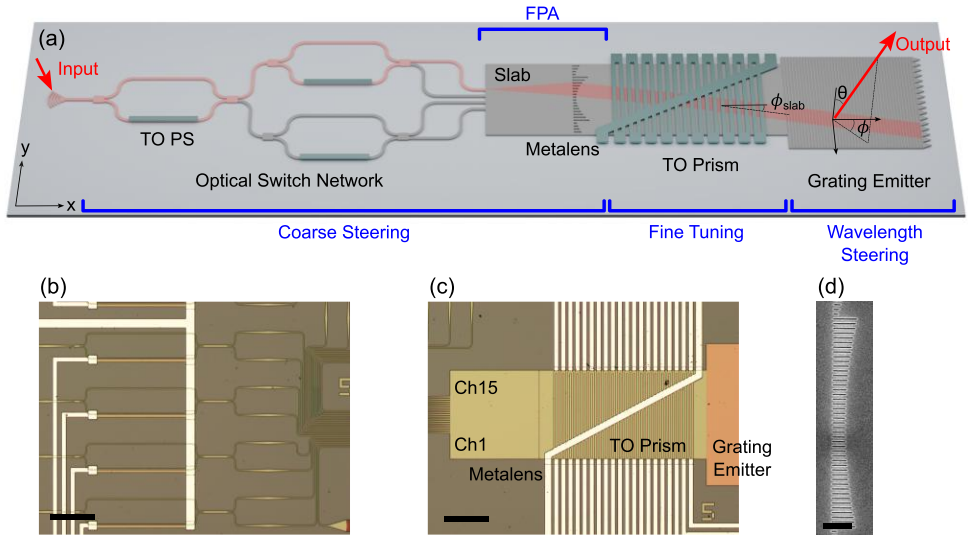

Fig. 1. (a) Schematic of the silicon photonic beam steerer. FPA: focal plane array. TO: thermo-optic. PS: phase shifter. Here, a simplified 1×4 optical switch network is plotted for clarity, while the actual device is 1×16. (b–c) Fabricated device under optical microscope, including (b) the optical switch network, and (c) metalens, TO prisms, grating emitter. Scale bar: 100 µm. Ch: channel. (d) SEM image of the on-chip metalens. Scale bar: 3 µm.

The schematic of the silicon photonic beam steerer is shown in Fig. 1(a). The device comprises an optical switch network, a metalens FPA, two TO prisms, and a shallow-etched slab grating [27]. Azimuthal ($\phi$) beam control is realized through discrete coarse steering by the metalens FPA and continuous fine-tuning by the TO prism. In the altitude ($\theta$) direction, wavelength-dependent steering is achieved through dispersive diffraction from the slab grating. Optical microscope (OM) images of the device fabricated on a silicon-on-insulator (SOI) wafer are shown in Fig. 1(b–c). The 1×16 optical switch network routes the input light to a selected waveguide channel that terminates at a slab waveguide. The FPA is formed by 15 waveguide channels (3 µm wide, arranged at a 5 µm pitch) and a metalens. The metalens, positioned at a focal length of 200 µm from the terminations, collimates light launched from each channel. The on-chip metalens, as shown by the scanning electron microscope (SEM) image in Fig. 1(d), is composed of subwavelength slots [30–32]. The metalens implements an optimized phase delay profile corresponding to an aberration-free F-theta lens [12]. When we launch light (in $TE_0$ mode) from one of the waveguide channels, the metalens collimates it into a well-defined propagation direction inside the slab. Channel selection determines the propagation direction [see the path in Fig. 1(a)], enabling discrete coarse steering across 15 distinct angles. To enable continuous fine-tuning of the beam direction, we implement microheaters—referred to as "thermo-optic prisms"—to locally heat one of the two triangular regions of the silicon slab waveguide. Owing to the large thermo-optic coefficient of silicon ($dn_{Si}/dT = 1.86 \times 10^{-4}$ K$^{-1}$), localized heating increases the refractive index, forming a thermo-activated prism that dynamically fine-tunes the guided beam direction. Finally, the guided wave within the slab is diffracted to the air by the shallow-etched slab grating, enabling beam emission with wavelength-dependent steering.

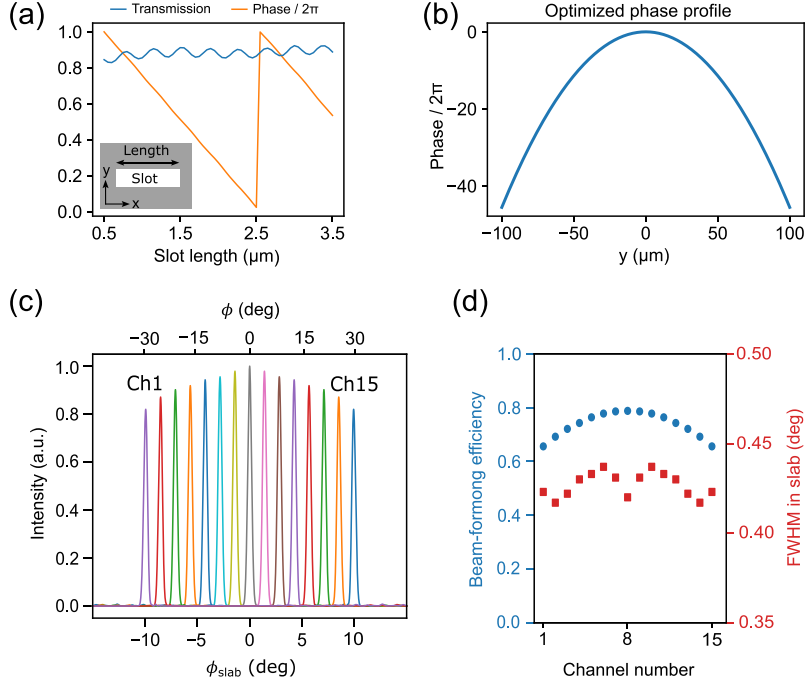

Fig.2. (a) Meta-atom phase and transmission library. The inset shows the unit cell of the metalens. (b) Optimized phase profile of the F-theta metalens. (c) FDTD-simulated angular distribution of the wave in the slab waveguide after passing the metalens. Different colors represent launching from different channels. (d) FWHM divergence angle inside the slab waveguide and beam-forming efficiency of the metalens.

We further elaborate on the beam-steering mechanisms enabled by each component. To enable active switching, we incorporate an optical switch network for channel selection. As shown in Fig. 1(a), the optical switch network consists of Mach-Zehnder switches arranged in a binary tree structure. The Mach-Zehnder arms are balanced to ensure broadband operation. A 1×$N$ optical switch network has $N-1$ switches in total, while only $\log_2 N$ switches need to be activated simultaneously to define the light path [12,25]. In our demonstration, only 4 switches in the 1×16 optical switch network are activated at any given time, enabling low power consumption for coarse steering. We use 15 out of the 16 output channels for subsequent light routing. To reduce propagation loss, we use 1.2-μm-wide multimode waveguides to route most of the optical paths within the network. Single-mode operation is maintained by employing adiabatic Euler bends [33].

We design an on-chip metalens to achieve aberration-free beam collimation, eliminating the need for packaging a bulky external lens. The metalens consists of subwavelength slots [30] with a period of 500 nm and a width of 140 nm. The slot length is varied in finite-difference time-domain (FDTD) simulations to construct a meta-atom phase library, as shown in Fig. 2(a). By varying the slot length from 500 nm to 2.5 μm, the library covers a full $2\pi$ phase range, allowing the creation of any desired phase profile. The phase profile of an F-theta lens is obtained through ray-tracing optimization (see Supplement 1) [12], and the optimized profile is shown in Fig. 2(b). According to the ray-tracing results, this design enables aberration-free collimation up to a propagation angle of $\pm 9.9°$ within the silicon slab. We further validate the metalens FPA design using 3D full-wave FDTD simulation. The far-field angular distribution of the 15 channels is shown in Fig. 2(c), plotted as a function of $\phi_{\text{slab}}$, the propagation angle inside the silicon slab. The FDTD simulation confirms well-collimated beams up to $\phi_{\text{slab}} = \pm 9.9°$, in agreement with the ray-tracing design. Owing to the large refractive index contrast between silicon and air, the steerable angles in air after diffraction by the grating emitter are significantly larger than the $\pm 9.9°$ inside the slab, as discussed later. Figure 2(d) shows the simulated divergence angle and the beam-forming efficiency (defined in Supplement 1). The average divergence angle inside the slab waveguide is 0.43°, corresponding to 1.22° in air. The beam-forming efficiency reaches 79 % at the central channel (Ch 8). Note that this efficiency reflects only the metalens' performance in forming collimated beams within the slab and does not include the grating emitter efficiency.

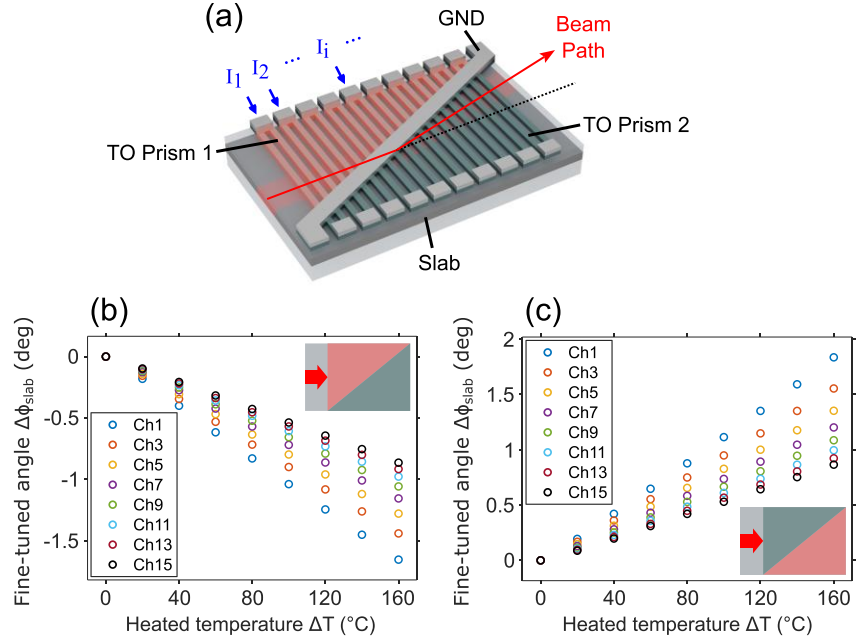

Fig.3. (a) Schematic of the TO prisms, which allow independent heating of two triangular regions of the slab waveguide to introduce fine beam deflection. The schematic is simplified for clarity, while in the actual device, each TO prism has 20 densely-packed microheater sections. GND: ground. (b–c) Simulated fine-tuning of the beam direction at different elevated temperatures with TO prism 1 (b) and TO prism 2 (c).

To enable continuous steering in the $\phi$ direction, we implement a fine-tuning mechanism based on a pair of TO prisms formed by tungsten microheaters above the silicon slab waveguide, as shown in Fig. 3(a). When current is applied, the microheaters locally heat a triangular region of silicon. Owing to the thermo-optic effect, the heated region exhibits an increased refractive index and functions as a dynamic prism that deflects the guided beam. Similar concepts of thermo-activated components have been demonstrated in other integrated photonic platforms [34]. The fine-tuning mechanism introduced here closes the blind spots of the FPA-based steering and increases the number of resolvable points. The same current is applied through 20 densely packed microheater sections (i.e., $I = I_1 = I_2 = I_3 \cdots$). The microheater sections are equally spaced and have the same width. Because the resistance of each section ($R_i$) is proportional to its length ($l_i$), the dissipated heat per unit length ($I^2 R_i / l_i$) is constant for all sections. This design creates uniform heating across the triangular region (see Supplement 1 for detailed geometry). The two TO prisms, activated by the upper half and lower half of the microheater sections, respectively, share a common ground and can be independently activated. We then perform FDTD simulations to estimate the fine-tuning angle as a function of the elevated temperature. To reduce computational cost, a smaller triangular region with the same slope is simulated (see Supplement 1). The simulated fine-tuned propagation angle inside the slab waveguide as a function of the elevated temperature $\Delta T$ is shown in Fig. 3(b) and (c). The beam can be fine-tuned in either the positive or negative direction, depending on which triangular region is activated. The angle change of the central beam exceeds its full width at half maximum (FWHM) when the elevated temperature is greater than approximately 60 °C.

We employ a shallow-etched grating emitter to diffract light from the guided slab mode into air and enable wavelength-dependent steering in the altitude ($\theta$) direction. The relationships

between diffraction angles in air and the propagation angles inside the slab waveguide are given by

$$k_0 \sin \phi = n_{slab} k_0 \sin \phi_{slab} \quad (1)$$

$$k_0 \sin \theta = n_{slab} k_0 \cos \phi_{slab} - \frac{2\pi}{\Lambda} \quad (2)$$

Equation (1) represents Snell's law. Here, $n_{slab}$ and $\Lambda$ denote the effective refractive index of the slab mode and the grating period, respectively. $\theta$ and $\phi$ are the altitude and azimuthal angles in air [defined in Fig. 1(a)]. Equation (1) is used to convert $\phi_{slab}$ to $\phi$ to predict the emission azimuthal angle in air, as shown on the top horizontal axis of Fig. 2(c). In our architecture, $\phi_{slab}$ is actively controlled by the FPA and TO prism. Owing to the large refractive index contrast between silicon and air, $\phi$ is significantly larger than $\phi_{slab}$, enabling a large azimuthal FOV in air of up to $\pm 29.3°$ using coarse steering alone [see Fig 2(c)]. The grating emitter has a period $\Lambda$ = 580 nm and a duty cycle of 0.25, with a shallow etch depth of 20 nm to enable a large emission aperture. An anti-reflection subwavelength structure is incorporated at the end of the grating emitter, as shown in Fig. 1(a). The emission $\theta$ angle at 1550 nm is designed to be 8.75°. The altitude FOV is determined by the grating dispersion and the laser tuning range.

## 3. Device fabrication, packaging, and electrical driving

We fabricate the beam steerer on a SOI wafer with a 220-nm top silicon layer and a 3-µm buried oxide (BOX) layer. Electron-beam lithography (VOYAGER, Raith) is first used to define the fully etched waveguide layer with a negative photoresist (ma-N 2403). The pattern is transferred into silicon via plasma etching (Plasma Pro 100 Estrelas, Oxford Instruments). A similar procedure is then performed using a positive photoresist (ZEP520A) to create the shallow-etched grating. A 1.3-µm-thick silicon dioxide cladding is subsequently deposited by plasma-enhanced chemical vapor deposition (PECVD) (Plasmalab 80 Plus, Oxford Instruments). On top of the cladding, we further sputter Ti/W with a total thickness of 143 nm and use lift-off to create the microheater layer. Each TO phase shifter, formed by this Ti/W layer, has a width of 8 µm and a length of 200 µm. Finally, a 1-µm-thick aluminum layer is deposited by thermal evaporation and patterned via lift-off to form the wires for electrical connection.

The fabricated chip is diced and mounted onto an aluminum holder using silver paste. We wire-bond the chip to printed circuit boards (PCBs), which interface with a multichannel programmable voltage source (PXIe-6739, National Instruments) for electrical driving, as shown in Fig. 4(a). This voltage source provides 64 independent channels with an update rate of 350 kS/s, enabling control of 15 TO phase shifters in the optical switch network and 40 microheater sections in the TO prisms. For the TO prisms, we first measure the resistance of each section and then apply the corresponding driving voltage to generate a uniform temperature distribution across the triangular prism region.

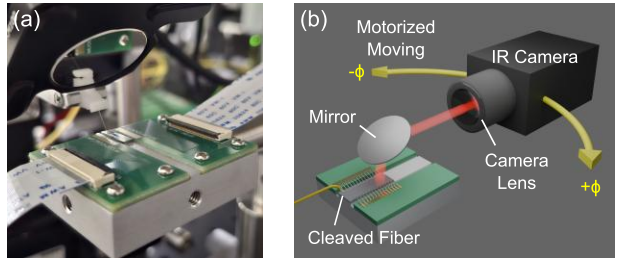

Fig. 4. (a) Photography of the packaged chip. The silicon photonic chip is diced to a size of 10.5 mm × 3 mm. (b) Schematic of the setup for beam characterization.

## 4. Beam-steering characterization

We experimentally demonstrate continuous beam steering over an azimuthal FOV of 62°. We characterize the beam-steering performance by constructing a Fourier imaging system, as illustrated in Fig. 4(b). A camera lens maps the 2D far-field angular distribution onto an infrared camera (Bobcat 320, Xenics). We precisely rotate the system in the azimuthal direction using a motorized stage to allow measurements over a wide FOV. Light from a tunable laser (CTL 1550, Toptica) is coupled into the input grating coupler via a cleaved single-mode fiber.

Figure 5(a) demonstrates the measured coarse steering in the $\phi$ direction and wavelength steering in the $\theta$ direction. The FPA-based coarse steering generates 15 discrete beams controlled by the optical switch network. The wavelength is fixed at 1550 nm in this measurement. This figure also shows the dispersive beam steering in the $\theta$ direction as the wavelength is tuned to 1540, 1560, 1570, 1580, and 1590 nm. These wavelengths enable an altitude FOV of 8.3°, currently limited by the angular measurement range of the Fourier imaging system. The measured angular dispersion of the grating emitter is 0.17°/nm. In Fig. 5(b), we combine both FPA-based coarse steering and TO-prism-based fine-tuning to show continuous steering over an FOV of 62°. To characterize the fine-tuning mechanism, we apply 1.3 W of electrical power to one of the TO prisms and measure the beam displacement of different channels, as shown in Fig. 5(c). The fine-tuned steering angle with Channel 1 reaches 2.47°, corresponding to 168% of the beam's FWHM. For Channel 15, a smaller tuning angle of 1.35° is observed, corresponding to 91% of the beam's FWHM. The difference in tunability between channels is also observed in the simulation (see Fig. 3). Figure 5(c) further shows that the combined operation of coarse steering and fine-tuning enables at least 45 resolvable points using only 15 waveguide channels. Since conventional OPAs and FPAs are typically constrained to a number of resolvable points comparable to the channel count [12], this result represents a three-fold enhancement in steering resolution. Figure 5(d) presents the fine-tuning of the central beam (Ch 8) under different applied electrical powers to the TO prism, demonstrating continuous operation. Figure 5(e) presents the measured SLSR and FWHM divergence angles for different channels without fine-tuning. The central beam (Ch 8) exhibits FWHM divergence angles of 1.45° in $\phi$ and 0.31° in $\theta$. The measured average FWHM divergence angle in $\phi$ is 1.48°, slightly larger than the simulated value of 1.22°. The beam quality remains high, as evidenced by the average SLSR of 19 dB.

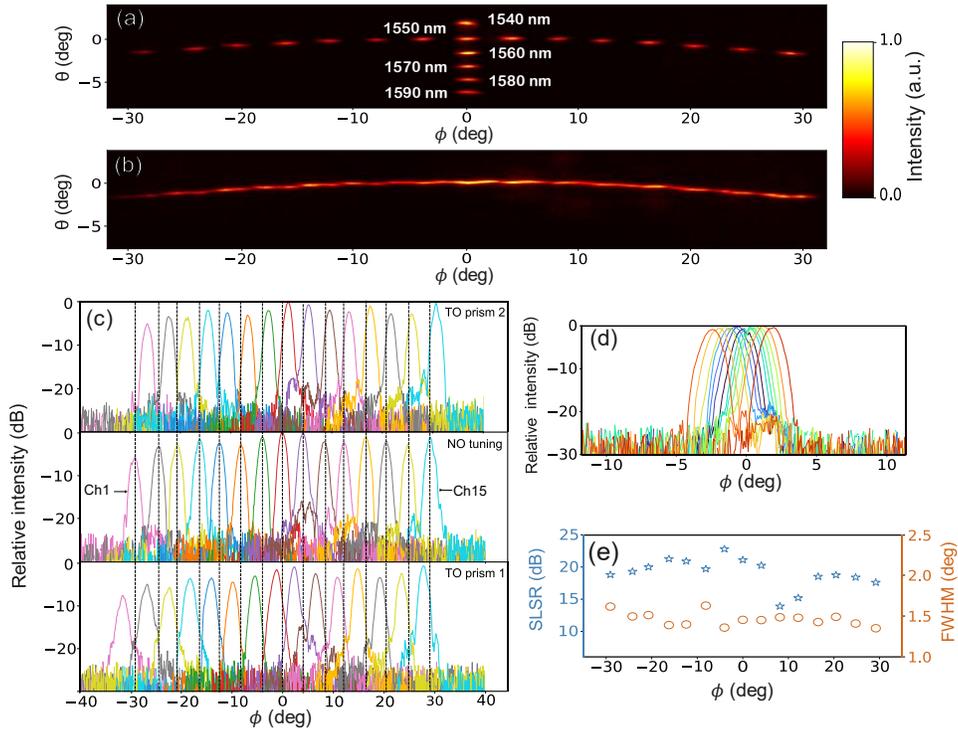

Fig. 5. (a) Far-field steering map demonstrating FPA-based coarse beam steering with 15 channels and wavelength tuning. The map is produced by overlapping the images measured from 15 channels and 6 wavelengths. (b) Far-field steering map combining FPA-based coarse steering and TO-prism-based fine-tuning, showing continuous steering over a 62° FOV. The map is produced by overlapping the images measured from 15 different channels and 3 different TO prism settings (1.3 W applied to TO prism 1, no fine-tuning, and 1.3 W applied to TO prism 2). (c) Azimuthal cross-section of the far-field steering map showing both coarse steering and fine-tuning. Different colors denote different channels. From top to bottom: 1.3 W applied to TO prism 2, no fine-tuning, and 1.3 W applied to TO prism 1. (d) Fine-tuning of the central beam under different applied electrical power to the TO prism. Power steps of 1.30, 0.92, 0.68, 0.48, 0.30, and 0.17 W are applied. (e) The measured SLSR and FWHM divergence angles for different channels.

We demonstrate programmable active beam steering and characterize the steering speed, achieving time constants on the order of 10 µs for both coarse steering and fine-tuning mechanisms. By programming the multichannel voltage source and the tunable laser, the output beam can be steered along a predetermined trajectory. As an illustrative example, Figure 6(a) shows a steering map in which the beam traces the letter "H" in the far field. To characterize the steering speed, we aim the beam at a photodetector (PDA20CS2, Thorlabs) and modulate the beam direction using coarse steering and fine-tuning mechanisms, respectively, while recording the photodetector response. The coarse steering speed is determined by the reconfiguration time of the optical switch network, governed by the TO phase shifters. As shown in Fig. 6(b), we reconfigure the optical switch network to move the beam into and out of the photodetector. The measured rise (heating) time, defined as the time from 10% to 90% of the signal amplitude, is 17.0 µs, while the fall (cooling) time is 8.8 µs. For fine-tuning speed characterization, we first align the beam to the photodetector and then toggle the TO prism on and off. The result is shown in Fig. 6(c). The measured fall time (heating of the TO prism) is

15.1 µs, and the rise time (cooling) is 22.3 µs. Notably, despite the substantially larger heated area of the TO prism, its response speed remains comparable to that of conventional TO phase shifters.

We further evaluate the power consumption and insertion loss of the silicon photonic beam steerer. The $P_{2\pi}$ of each TO phase shifter, defined as the electrical power required to induce a $2\pi$ phase shift, is 40 mW, a typical value for the silicon photonic platform [13]. The maximum power consumption of the optical switch network, consumed by $\log_2 N$ switches, is $4 \times P_{2\pi} = 160$ mW. The measured extinction ratio of a single optical switch is 18.6 dB. The TO prism consumes up to 1.3 W of electrical power. As a result, the total electrical power consumption of the beam steerer, including both the optical switch network and the TO prism, has a maximum value of 1.46 W. The measured total insertion loss is 16.5 dB.

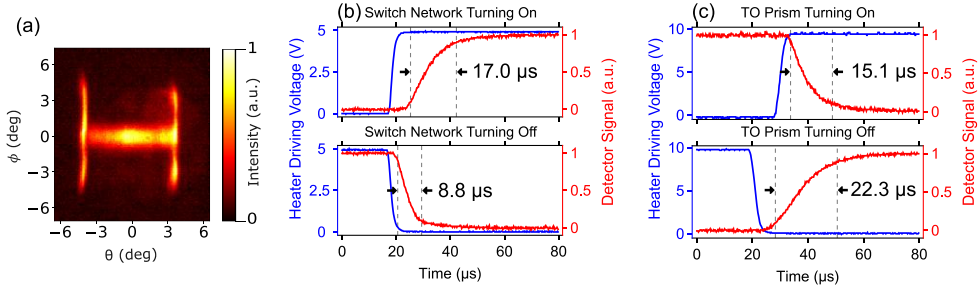

Fig. 6. (a) Steering map that overlaps the scans with a trajectory tracing the letter "H". (b) Coarse steering speed measurement. (c) Fine-tuning speed measurement.

## 5. Discussions

We have demonstrated a silicon photonic beam steerer based on a novel FPA architecture. An on-chip metalens eliminates the need for packaging a bulky external collimation lens. The incorporation of TO prisms removes blind spots, enables continuous fine-tuning, and enhances the number of resolvable points by a factor of three. Continuous beam steering is particularly critical for FSO links, where fine alignment and tracking between transmitters and receivers are required [35]. High beam quality is maintained across the FOV, as evidenced by an average SLSR of 19 dB. The architecture enables higher beam quality than conventional non-apodized OPAs, which have a theoretical maximum SLSR of 13 dB [16]. This architecture also offers more robust calibration compared to OPAs, as phase calibration is not required. The measured azimuthal FOV reaches 62°, currently limited by the aberration-free numerical aperture of the metalens, which can be further increased using doublet designs [36] or optimization algorithms [37]. We demonstrate 45 resolvable points with a relatively small array of only 15 waveguide channels. The architecture is readily scalable to a larger number of resolvable points by proportionally increasing the channel count $N$ together with the metalens focal length $f$. These results establish a promising route toward ultracompact, fully integrated beam steerers for next-generation LiDAR and FSO systems.

**Funding.** National Science and Technology Council (113-2622-8-A49-013-SB); National Science and Technology Council (111-2221-E-A49-014-MY3); National Science and Technology Council (114-2221-E-A49-039-MY3).

**Acknowledgments.** This research is partially supported by the Yushan Fellow Program by the Ministry of Education (MOE), Taiwan. The authors would like to acknowledge chip fabrication support provided by Taiwan Semiconductor Research Institute (TSRI), Taiwan. We thank the National Center for High-performance Computing (NCHC) for providing computational and storage resources.

# SILICON PHOTONIC BEAM STEERER BASED ON METALENS FOCAL PLANE ARRAY: SUPPLEMENTAL DOCUMENT

### 1. Design of the phase profile of the F-theta metalens

We use ray-tracing optimization to obtain the phase profile of an aberration-free F-theta metalens [1]. Prior to the ray-tracing design, we use 3D finite-difference time-domain (FDTD) simulation to calculate the mode expansion behavior from a 3 µm-wide waveguide to a large silicon slab waveguide. The equivalent numerical aperture (NA) is calculated to be 0.17 by fitting the divergent angle of the beam. Next, we use ray-tracing software and place point sources on the emission plane at 0 µm, 10 µm, 20 µm, and 30 µm from the optical axis to perform the optimization. As plotted in Fig. 2(b) of the main text, the optimized phase profile is given by

$$\varphi(r) = -7171(r/R)^2 + 2070(r/R)^4 - 1.09\times10^5(r/R)^6 \\ + 2.03\times10^6(r/R)^8 - 8.53\times10^6(r/R)^{10}, \tag{S1}$$

where $r$ is the distance from the center of the metalens, and $R$ is a normalized radius that equals 0.5 mm. This phase profile, according to the ray-tracing results, allows aberration-free beam collimation for the focal plane array (FPA) up to an output beam propagation angle of $\pm 9.9°$ in the silicon slab.

### 2. Calculation of the beam-forming efficiency of the metalens

We further verify the metalens FPA design by performing a two-stage 3D FDTD simulation. In the first stage, we simulate the expanding profile of the $TE_0$ mode from a 3 µm-wide waveguide channel into the silicon slab and record the transverse electromagnetic fields after a propagation distance of 200 µm. In the second stage, the recorded fields are used as the source to the metalens, and far-field projection is performed after the light passes through the metalens. The far-field angular distribution from each channel is consistent with the ray-tracing design, as shown in Fig. 2(c) of the main text. To quantify the performance of the metalens, we define the beam-forming efficiency as follows:

$$\text{Beam-forming efficiency} = \frac{\int_{\phi_{max}-3FWHM}^{\phi_{max}+3FWHM} I(\phi)d\phi}{P_{in}}. \tag{S2}$$

$\phi_{max}$ is the angle at the peak of a specific channel. $FWHM$ is the full width at half maximum divergence angle of the channel. $I(\phi)$ is the intensity. $P_{in}$ is the input power to the metalens. The beam-forming efficiency indicates how much of the input light entering the metalens contributes to the collimated beam within $\pm 3\ FWHM$ range.

### 3. Detailed microheater geometry of the thermo-optic prisms

The microheater geometry of the thermo-optic (TO) prisms is shown in Fig. S1. It consists of 20 sub-heaters and a 15 µm-wide shared ground. Each sub-heater consists of two 4-µm-wide narrow heaters separated by 4 µm. The heated area is a triangle with a base of 331.4 µm and a height of 174.9 µm. The resistance is designed to be approximately 1 kΩ, allowing it to be driven by the multichannel programmable voltage source (PXIe-6739, National Instruments). Next, we place 20 identical sub-heaters in an array with a period of 16 µm, together with a shared ground along a diagonal from the lower left to the upper right, separating the structure into two triangles. This configuration forms two independently controllable TO prisms.

To operate the device, we apply the same current to each sub-heater to ensure uniform temperature across the entire region. At any given time, only one side is powered while the other side remains unpowered. As a result, a uniform temperature distribution is formed in the triangular region. The current is collected by the shared ground at the center.

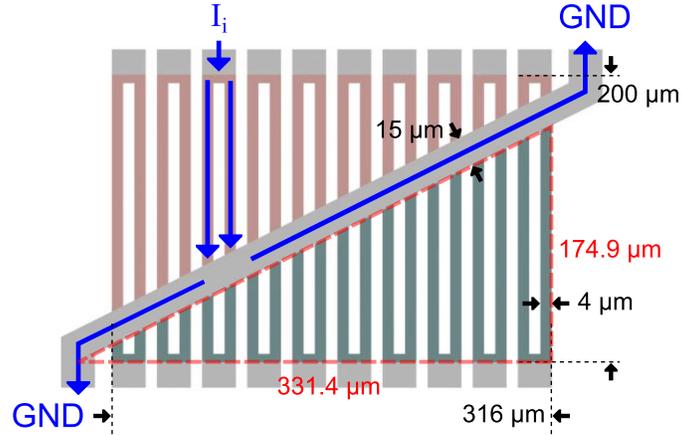

Fig. S1. Illustration of the thermo-optic prisms in top view. The blue arrows indicate the current flow. We show only 20 channels in this illustration for simplicity, while the actual device has 40 channels. GND: ground.

## 4. Simulation of the beam deflection by the thermo-optic prism

The schematic of the simulation model is shown in Fig. S2, where the slab waveguide has a triangular region with a uniformly increased refractive index based on the thermal-optic coefficient of silicon. We use 2.5D FDTD to simulate a smaller triangular region (189 μm × 100 μm) with the same slope of 27.8° to save computational resources. Since the model is smaller than the actual TO prism, we launch a scaled Gaussian beam with a beam waist of 17.5 μm into a 100 μm × 220 nm silicon slab waveguide. The channel spacing is 2.5 μm in the y-axis. The far-field angle is calculated by capturing the electromagnetic fields after the heated triangular region.

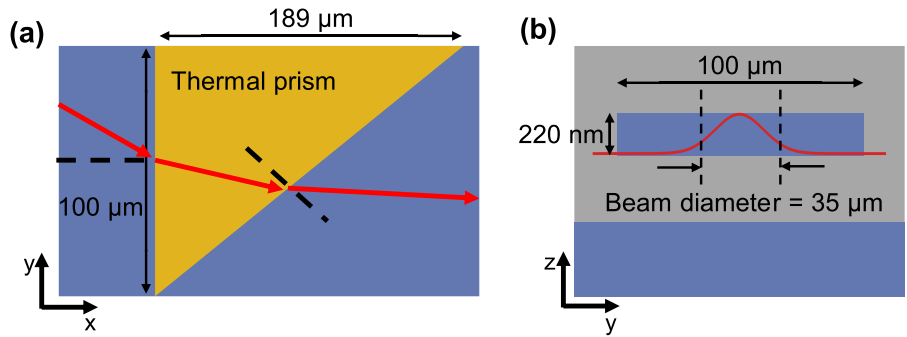

Fig. S2. Simulation model of the thermo-optic prisms in the (a) longitudinal and (b) transverse cross-sections.